\documentclass[12pt]{article}

\usepackage{graphicx}
\usepackage{amssymb}
\usepackage{amsmath}

\numberwithin{equation}{section}

\begin{document}
\baselineskip=16pt
\begin{titlepage}

\begin{center}
\vspace*{12mm}

{\Large\bf%
Negative modes of Schwarzschild black hole in Einstein-Gauss-Bonnet Theory%
}\vspace*{10mm}

Takayuki Hirayama
\vspace*{4mm}

{\it Physics Division, National Center for Theoretical Sciences, Hsinchu 300, Taiwan 
\\
and
\\
Department of Physics, National Taiwan Normal University, Taipei 116, Taiwan
}\\[1mm]

\end{center}
\vspace*{10mm}

\begin{abstract}\noindent%
We study non conformal negative modes of Schwarzschild black holes in the five dimensional Einstein-Gauss-Bonnet gravity theory with a zero or negative cosmological constant, in the context of the semiclassical quantum gravity formulated by the path integral. 
We find an unique negative mode when the black hole has a negative heat capacity which is the same in Einstein theory. On the other hand, we still find negative modes inside a parameter region where a small black hole has a positive heat capacity. 
The number is one/two for the case of zero/negative cosmological constant respectively. In the rest of the parameter region where the heat capacity is positive, we find no negative modes. We discuss the possible physical understanding of having one or two negative modes.
\end{abstract}

\end{titlepage}

\newpage

\section{Introduction}

Black holes are very important objects in classical and quantum gravity. One of the most important and profound properties in black hole physics is its thermodynamics.
The thermodynamic quantities can be derived from the path integral formulation of quantum gravity, and thus have connections to quantum nature of gravity.

In the case of four dimensional Euclidean Einstein quantum gravity, the Schwarzschild black hole is an instanton to induce a decay of hot flat space by nucleation of black holes. 
This instability is suggested by a unique off-shell negative mode around the Schwarzschild instanton~\cite{Gross:1982cv}. The existence of negative mode is believed to have a connection with the local thermodynamic stability of Schwarzschild black hole. For example, the Anti de-Sitter (AdS) Schwarzschild black hole has an unique negative mode when the black hole has a negative heat capacity, and the negative mode disappears exactly when the black hole becomes local thermodynamic stable~\cite{Prestidge:1999uq}. Reall~\cite{Reall:2001ag} discussed in detail the connection between the local thermodynamic stability and the stability against small perturbations around a black hole solution (existence of negative modes) and showed they agree under some reasonable assumptions.

The negative mode has been also studied in the connection with Gregory-Laflamme instability~\cite{Gregory:1993vy}, since the negative eigen value is related with a Kaluza-Klein mass square of graviton along the directions to which a black string extends~\cite{Reall:2001ag}. Gubser and Mitra~\cite{Gubser:2000ec} conjectured that there is a negative mode iff the black hole has a thermodynamic instability (the correlated stability conjecture). This conjecture is known to be satisfied for many different black branes including the  black branes discussed by Gregory and Laflamme~\cite{Gregory:1993vy}, black p-brane solutions in string theory~\cite{Hirayama:2002hn} and also black strings in Anti de-Sitter space~\cite{Prestidge:1999uq, Gubser:2000ec, Hirayama:2001bi}. 
Also more complicated systems such as the D0-D2 bound state~\cite{Gubser:2004dr} and the non-extremal smeared black branes~\cite{Harmark:2005jk} are discussed and the conjecture is satisfied.
There are, on the other hand, counter examples~\cite{Friess:2005zp}, and those are related with the non uniqueness of black string under fixed conserved charges. The correlated stability conjecture is checked to be true for all the known cases when a black string is uniquely determined by the conserved charges.

Since the black hole thermodynamics has a deep connection with quantum mechanics of gravity, we have a natural question whether the properties related with the black hole thermodynamics are modified, if we add higher dimensional corrections to Einstein Hilbert action which are expected to appear in an effective theory of quantum gravity, for example, string theory. 
The Gauss-Bonnet term is one of such terms and is the lowest dimensional term among them. The black hole solutions are known~\cite{Boulware:1985wk} and the thermodynamics is also calculated~\cite{Myers:1988ze} in Einstein-Gauss-Bonnet theory. 
We naively expect that the local thermodynamic stability and the stability against small perturbations still agrees in Einstein-Gauss-Bonnet theory. In this paper we study the negative mode of Schwarzschild black hole in Gauss-Bonnet theory with and without a negative cosmological constant. Since the Gauss-Bonnet term becomes topological in four dimensions, we study the five dimensions. 

We find there is an unique negative mode when the black hole has a negative heat capacity. As the black hole becomes larger, the negative eigenvalue goes up and becomes zero exactly when the sign of heat capacity changes from negative to positive. This behavior is exactly the same as that in Einstein gravity and consistent with the fact that the effect from Gauss-Bonnet term is negligible when a black hole has a large size and the spacetime is weakly curved. However as the five dimensional black hole becomes smaller and starts having a positive heat capacity due to the Gauss-Bonnet term, we find a new behavior. We still find negative modes and the numbers of negative modes are one or two for the case of zero or negative cosmological constant respectively. We also find one or two positive modes for the case of negative cosmological constant. As the black hole becomes further smaller, the negative modes disappear.
These results lead us to reconsider the connection between the thermodynamic stability and the existence of negative modes.

\vspace{1ex}

In the next section, we review the thermodynamics of the Schwarzschild black hole in Gauss-Bonnet theory. Then in section 3, we carry on the S-mode perturbations around the black hole and study the non conformal negative modes. We only study a simple case, i.e. the five dimensional Gauss-Bonnet theory. The detail procedure of numerical analysis is summarized in Appendix A. We summarize our numerical results as figures in Fig.\ref{w1}-\ref{t3} (and also Fig.\ref{w4}-\ref{t4} in Appendix B). The real numerical data is listed in Appendix B. In section 4 we interpret our numerical results and conclude in section 5.


\section{Black hole thermodynamics}

In this section we summarize the setup and thermodynamic quantities of black hole  in Gauss-Bonnet theory.

The path integral formalism of Euclidean quantum gravity with a finite temperature $T$ is schematically written 
\begin{align}
  Z &=  \int \! {\cal D}g \exp[-I(g)],
  \hspace{3ex}
  F = -T \ln Z ,
\\
  I &=  \frac{1}{16\pi G_n}
  \int_M \! d^n x \sqrt{g}\left[ R -2\Lambda +\frac{c}{2} L_{GB} \right]
  +\frac{1}{8\pi G_n}\int_{\partial M}d^{n-1} \! \sqrt{h}K ,
\\
  L_{GB}  &=  R^2 -4 R_{ab} R^{ab} + R_{abcd} R^{abcd} ,
\end{align}
where $F$ is the free energy, $G_n$ is the Newton constant, $\Lambda$ is the cosmological constant, 
$c$ is a coupling constant of mass dimension $-2$ in front of Gauss-Bonnet term $L_{GB}$, $h$ is the induced metric on the boundary, $K$ is a surface term and $n$ is the spacetime dimensions. We also need gauge fixing terms. In this section we consider $\Lambda<0$. The manifold is asymptotically $S^1\times R^{n-1}$ and the periodicity for $S^1$ along the Euclidean time $\tau$ is given $\tau=\tau+\beta, \beta=1/T$.

The partition function $Z$ is semiclassically approximated by the sum of classical paths (saddle point approximation). We are then interested in a black hole solution and the small perturbations about the solution. The black hole solution in the $n$ dimensions is known~\cite{Boulware:1985wk}
\begin{align}
  ds^2 &=  \bar{g}_{ab} dx^a dx^b
  = f(r) d\tau^2 +\frac{dr^2}{f(r)} + r^2 d\Omega_{n-2}^2 , 
\\
 f(r) &= 1 
 +\frac{r^2}{c(n-3)(n-4)}\left\{ 1 +\epsilon\sqrt{1
 +\frac{4c(n-3)(n-4)}{(n-2)(n-1)} \left[
 \frac{(n-1)\mu}{r^{n-1}} +\Lambda \right]
  } \right\} ,
  \label{solf}
\end{align}
where $\mu$ is a parameter and is related with the black hole mass $M=\mu{\cal A}_{n-2}/(8\pi G_n)$ where ${\cal A}_{n-2}$ is the area of unit $(n-2)$ sphere. There are two solutions $\epsilon=\pm$ and the solution with $\epsilon=-1$ becomes the Schwarzschild black hole in Einstein gravity in the limit of $c\rightarrow 0$. Thus we only discuss the case of $\epsilon=-1$ in this paper. The temperature $T$ is determined to avoid a conical singularity at the horizon and then given by the surface gravitational force at the horizon $r=r_h$, ($f(r_h)=0$),
\begin{align}
  T =  \frac{f'(r_h)}{4\pi} =
  \frac{ (n-2)(n-3)\left\{2r_h^2 +c(n-4)(n-5)\right\} -4r_h^4\Lambda }
  {8\pi r_h(n-2) \left\{ r_h^2 +c(n-3)(n-4) \right\} } 
\end{align}
for $r_h^2>-c(n-3)(n-4)$. Using the horizon radius $r_h$, the black hole mass is written
\begin{align}
 M = \frac{{\cal A}_{n-2}}{8\pi G_n}
 \left[ \frac{n-2}{2}r_h^{n-3} +\frac{c(n-2)(n-3)(n-4)}{4}r_h^{n-5} 
 -\frac{\Lambda}{n-1} r_h^{n-1} \right] .
\end{align}
The entropy which satisfies the first law in thermodynamic mechanics is computed~\cite{Myers:1988ze} from $I(\bar{g}) = \frac{1}{T}M(\bar{g}) -S(\bar{g})$,
\begin{align}
  S =  \frac{{\cal A}_{n-2}}{4G_n} \left[ r_h^{n-2} + n(n-1)cr_h^{n-4} \right],
\end{align}
where the entropy is no longer simply proportional to the area of the horizon. The local stability as a thermodynamic system is determined from the sign of heat capacity $C_V$,
\begin{align}
  C_V = \frac{\partial M}{\partial T} .
\end{align}
If we denote a one parameter family of geometries which contains the black hole solution as $x$, the second derivative of the action along the path $x$ is
\begin{align}
  \left( \frac{\partial^2 I}{\partial x^2} \right)_T
   = \left( T \frac{d x}{d T} \right)^{-2} \frac{dM}{dT}   
 \hspace{3ex}
 {\rm at~} 
  \left( \frac{\partial I}{\partial x} \right)_T =0 .
\end{align}
Thus if the specific heat is negative, the perturbations around the black hole where 
$(\partial I/\partial x)_T =0$ must have at least one negative mode~\cite{Reall:2001ag}. This is a basic observation that the local thermodynamic stability is equivalent with the stability against the perturbations about the black hole solution in the semi classical quantum gravity.


\paragraph{Black hole in five dimensions}

\noindent\vspace*{1ex}

The heat capacity depends on various parameters. Hereafter we study $n=5$, since the five dimensions is the lowest dimension where the Gauss-Bonnet term non trivially contributes to the local dynamics of gravity.
In five dimensions ($n=5$), the mass is a monotonically increasing function in terms of $r_h$ and then the sign of heat capacity is equal to the sign of $dT/dr_h$. Since the temperature is
\begin{align}
 T &= \frac{ 3-r_h^2\Lambda }{ 6\pi ( r_h + (2c/r_h) )} ,
\end{align}
for $r_h^2>-c/2$,
we divide three regions (1) $c\leq 0$, (2) $0< c < \frac{1}{-6\Lambda}$ and (3) $c \geq \frac{1}{-6\Lambda}$ depending on the behavior of heat capacity in terms of the black hole radius $r_h$.
\begin{itemize}
\item[(1)]
{\boldmath $c\leq 0$} : 
The temperature behaves similar to the case of AdS Schwarzschild black hole in Einstein gravity, but goes to infinity at $r_h=\sqrt{-2c}$ before $r_h=0$. See Fig.\ref{w1}. The temperature takes the lowest value at $r_h=r_c$,
\begin{align}
 r_{c} &= \sqrt{ \frac{1}{-2\Lambda} \left[ 3+6c\Lambda
 +\sqrt{ 3(1+6c\Lambda)(3+2c\Lambda)}
 \right] } .
\label{cv}
\end{align}
Thus $C_V<0$ for $r_h<r_c$ and $C_V>0$ for $r_h>r_c$.

\item[(2)]
{\boldmath $0< c < \frac{1}{-6\Lambda}$} : 
The temperature is largely modified at a small black hole region and there is a region where there are three different black holes at a same temperature. See Fig.\ref{w2}.
 The temperature starts from zero and increases up to the black hole radius $r_{a}$, and starts decreasing until $r_h=r_c$. After that, the temperature increases again as the black hole becomes larger. $r_c$ is same as \eqref{cv} and $r_a$ is
\begin{align}
 r_a &= \sqrt{ \frac{1}{-2\Lambda} \left[ 3+6c\Lambda
 -\sqrt{ 3(1+6c\Lambda)(3+2c\Lambda)}
 \right] } ,
\end{align}
As $c\rightarrow \frac{1}{-6\Lambda}$, $r_a$ approaches to $r_c$, i.e. $r_a\rightarrow r_c$.
$r_c$ does not change much by changing $c$. $C_V<0$ for $r_a<r_h<r_c$, and $C_V>0$ for $r_h<r_a$ and $r_h>r_c$.

\item[(3)]
{\boldmath $\frac{1}{-6\Lambda} \leq c$} :
The region where the black hole has a negative heat capacity disappears and the black hole is always stable as a thermodynamic system. See Fig.\ref{w3}. $C_V>0$ for all~$r_h$.
\end{itemize}
The behavior of temperature of black hole in five dimensions is special and that in more than five dimensions is almost the same as that in Einstein gravity.


\section{S-mode perturbations and negative modes}

We have described the thermodynamic properties of black hole in Gauss-Bonnet theory. We now study small perturbations around the black hole solution in five dimensions. The action around the soliton up to the quadratic orders in the perturbations $\delta g_{ab}$, ($g_{ab}=\bar{g}_{ab}+\delta g_{ab}$), becomes
\begin{align}
 I &= I_0(\bar{g}) + I_2(\bar{g},\delta g) , \hspace{3ex}
 I_2(\bar{g},\delta g) = \int \! d^nx \
 \delta g_a^{~b} \Delta_{~b~d}^{a~c}(\bar{g}) \delta g_c^{~d} ,
 \label{lich}
\end{align}
where the operator $\Delta(\bar{g})$ is called the Euclidean Lichnerowicz operator in Einstein theory. Thus the one loop quantum collections around the solution are proportional to $\sqrt{\det \Delta(\bar{g}) }$ and if we have one or more negative eigenvalues in this operator $\Delta(\bar{g})$, the black hole solution is not a local minimum, but a saddle point. 

We in this paper study the time independent s-mode perturbations around the black hole solution. This is because this mode is known to induce the unique negative mode of Schwarzschild black hole in Einstein gravity and we expect that negative modes appear only in the s-mode perturbations. The time independent s-mode perturbations are given
\begin{align}
  \delta g_a^{~b} = & {\rm diag} ( H_{tt}(r), H_{rr}(r), K(r), K(r), K(r) )  ,
\end{align}
where the Lorentz indices are made up and down by the background metric $\bar{g}_{ab}$. In the case of Einstein gravity, these fluctuations have two independent modes and the one mode is a trace mode and the other mode is a traceless mode. The trace mode is related with a conformal rescaling and has a ghost like kinetic term.
The traceless mode gives the unique negative mode. In Gauss-Bonnet theory, there are also two modes and we study the mode which becomes a traceless mode in the limit to Einstein gravity. We name this mode as the 'traceless' mode even though it is no longer traceless and study the negative eigenmodes in this mode. Similarly we name the other mode which becomes a trace mode in the limit to Einstein gravity as the 'trace' mode.

In order to compute the eigenvalues of the operator, we have to solve the equation of motion for the eigenmodes
\begin{align}
 \Delta_{~b~d}^{a~c}(\bar{g}) \delta g_c^{~d} &= \lambda \delta g^a_{~b} ,
\end{align}
with appropriate boundary conditions at the infinity and the horizon. We note that a eigenmode with $\lambda\neq 0$ is an off-shell mode and a mode with $\lambda =0$ is an on-shell classical perturbation. This eigen equation is complicated and we do not separate the two modes in the equation level, but we put the 'trace' mode zero by giving the appropriate boundary condition in the numerical analysis. We give the detail procedure of numerical analysis in Appendix A.


\paragraph{Asymptotic solutions}

\noindent\vspace*{1ex}

We first study the asymptotic behaviors of s-mode perturbations at the horizon and the infinity in $r$. The asymptotic solutions of the 'traceless' mode at the horizon are ($\epsilon=r-r_h$)
\begin{align}
  H_{tt}(r) &\propto 1 + {\cal O}(\epsilon) , &
  H_{rr}(r) &\propto -1 + {\cal O}(\epsilon), &
  K_{tt}(r) &\propto  \frac{2}{3} + {\cal O}(\epsilon).
  \label{ha}
\\
  H_{tt}(r) &\propto \frac{1}{\epsilon} + {\cal O}(\ln\epsilon) , &
  H_{rr}(r) &\propto -\frac{1}{\epsilon} + {\cal O}(\ln\epsilon), &
  K_{tt}(r) &\propto \frac{2}{3\epsilon} + {\cal O}(\ln\epsilon).
\label{ha2}
\end{align}
At the leading order, the asymptotic solutions are the same as those in Einstein theory and they are traceless. The first solution is a normalizable mode which becomes clear if we use Kruskal coordinate. 

Similarly the asymptotic solutions for $\Lambda\neq 0$ at the infinity are
\begin{align}
  H_{tt}(r) &\propto r^{a_++2} + {\cal O}(r^{a_+}) , &
  H_{rr}(r) &\propto r^{a_+} + {\cal O}(r^{a_+-2}), &
  K_{tt}(r) &\propto \frac{1}{3}r^{a_++2} + {\cal O}(r^{a_+}).
 \\
  H_{tt}(r) &\propto r^{a_-+2} + {\cal O}(r^{a_-}) , &
  H_{rr}(r) &\propto r^{a_-} + {\cal O}(r^{a_--2}), &
  K_{tt}(r) &\propto \frac{1}{3}r^{a_-+2} + {\cal O}(r^{a_-})  .
 \label{ifa}
\end{align}
Here
\begin{align}
  a_\pm = & -4 \pm 2
  \sqrt{1-\frac{\lambda}{(1-c\frac{-\tilde{\Lambda}}{3})\frac{-\tilde{\Lambda}}{3}}} ,
  \hspace{3ex}
  \tilde{\Lambda} = \frac{3}{c}\left[ \sqrt{1+\frac{2c}{3}\Lambda} -1
  \right] ,
\end{align}
where $\tilde{\Lambda}$ is an effective cosmological constant computed from the AdS black hole metric in the infinity, i.e.
\begin{align}
 f(r) &= 1 
 +\frac{r^2}{2c}\left\{ 1 -\sqrt{1
 +\frac{2c}{3} \left[
 \frac{4\mu}{r^4} +\Lambda \right]
  } \right\} 
  = 1-\frac{\tilde\Lambda}{6}r^2 +{\cal O}(r^{-2}) .
\end{align}
The asymptotic solutions are traceless at the leading order and the second solution \eqref{ifa} is a normalizable mode. 

\vspace{1ex}

Therefore a small perturbation must have the asymptotic solutions \eqref{ha} and \eqref{ifa} at the horizon and infinity. In order to obtain the eigenvalues and eigenfunctions, we rely on a numerical analysis. In the numerical analysis, we start from a point near the horizon with the normalizable mode (\ref{ha}), and solve the equations of motion toward the infinity. We change the value of $\lambda$ and find the eigenmode by the shooting method. See more detail in Appendix~A.


\paragraph{Negative modes}

\noindent\vspace*{1ex}

We now show our numerical results of eigenvalue $\lambda$. Here we only describe the results, and in the next section we will discuss how to understand the results.

As discussed before, the heat capacity depends on the parameter $c$ in front of Gauss-Bonnet term and we divide three regions depending on how the heat capacity behaves: (1) $c\leq 0$, (2) $0<c<\frac{1}{-6\Lambda}$ and (3) $c\geq \frac{1}{-6\Lambda}$. Since the equations of motion are invariant under the following rescaling
\begin{align}
 r &\rightarrow \alpha r,
 &
 M &\rightarrow \alpha^2 M,
 &
 c &\rightarrow \alpha^2 c,
 &
 \Lambda &\rightarrow \alpha^{-2} \Lambda,
 &
 \lambda &\rightarrow \alpha^{-2} \lambda,
\label{resc}
\end{align}
with $\alpha >0$, we can always take $\Lambda=-1$ in the numerical analysis.
The numerical data used in Fig.\ref{w1}-\ref{t3} are given in Appendix B.

We show our numerical result in Fig.\ref{w1} with $(c,\Lambda)=(-0.1,-1)$ in region (1).
\begin{figure}[t]
\begin{center}
\begin{minipage}{7.5cm}
\begin{center}
\includegraphics[height=7cm]{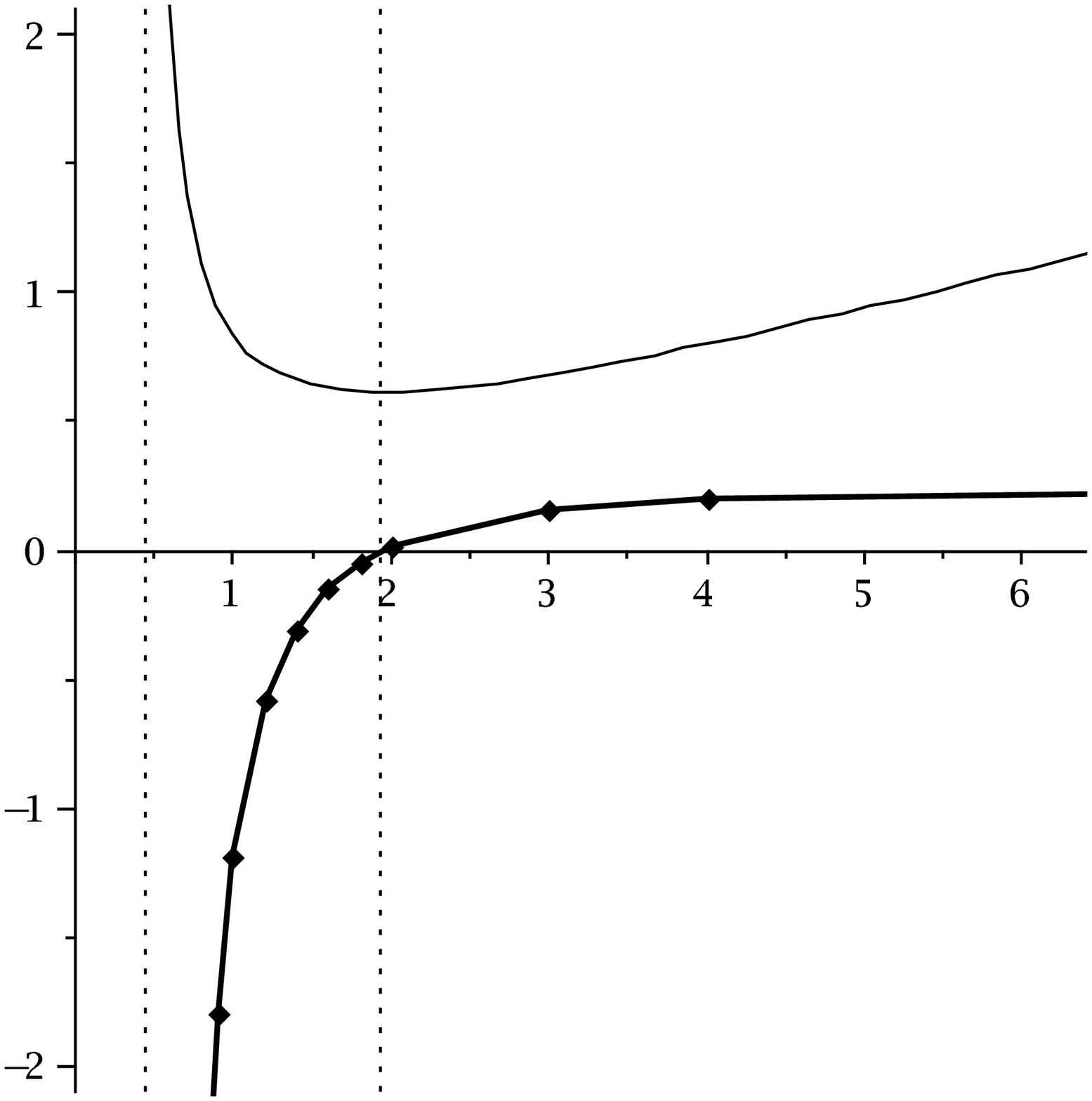}
\put(-5,90){$r_h$}
\put(-90,140){$T$}
\put(-145,25){$\lambda$}
\caption{The eigenvalue $\lambda$ in terms of $r_h$ for $c=-0.1$ with $\Lambda=-1$. The value on the vertical line is for $\lambda$ (not for $T$). 
$r_c=1.94$. $\lambda\rightarrow -\infty$,  ($T\rightarrow\infty$), at $r_h=\sqrt{-2c}$.}
\label{w1}
\end{center}
\end{minipage}
\hspace{4ex}
\begin{minipage}{7.5cm}
\begin{center}\vspace{-5ex}
\includegraphics[height=7cm]{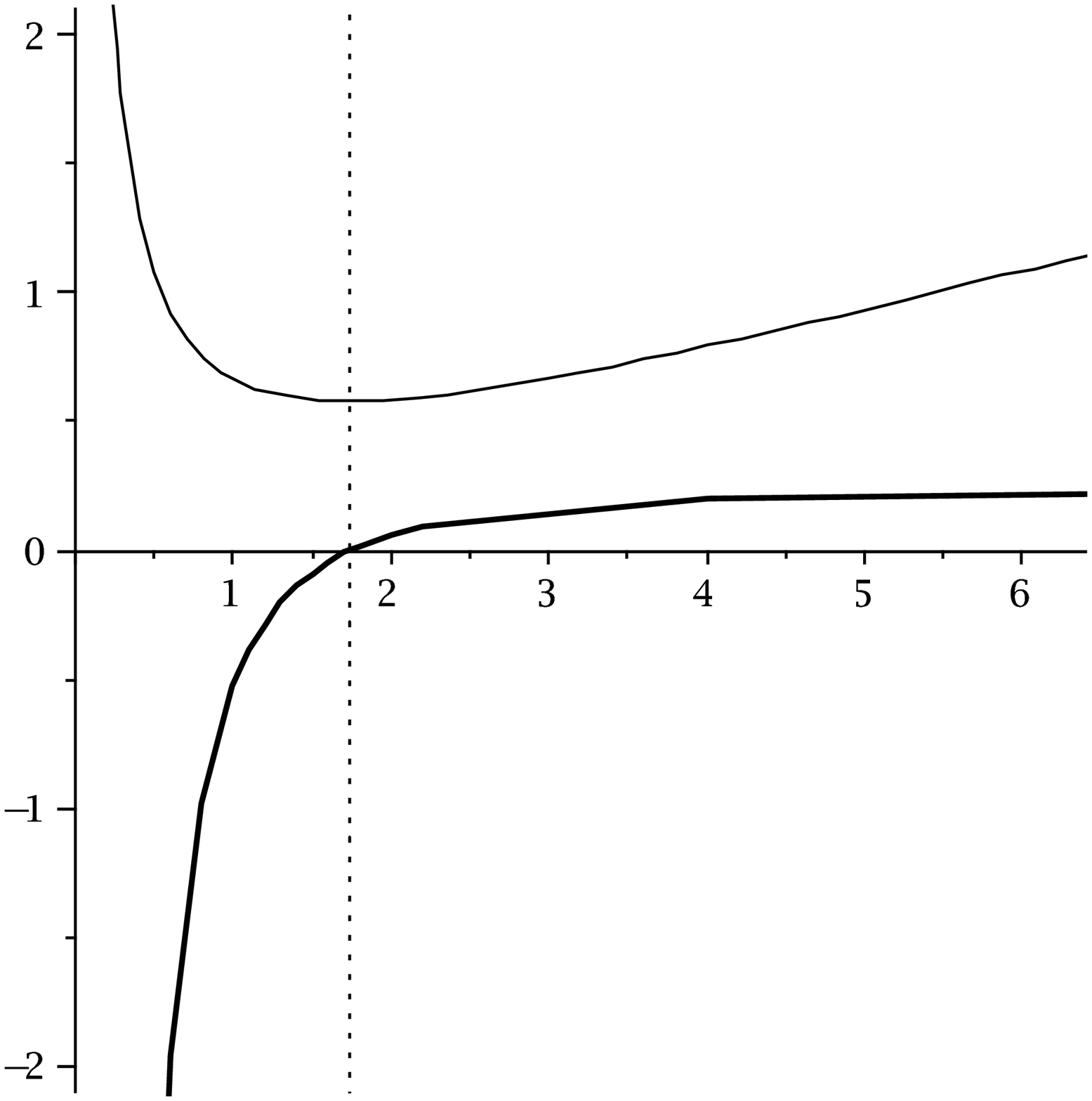}
\put(-5,90){$r_h$}
\put(-90,140){$T$}
\put(-155,25){$\lambda$}
\caption{The eigenvalue $\lambda$ in terms of $r_h$ for $c=0$ (Einstein theory) with $\Lambda=-1$.}
\label{t1}
\end{center}
\end{minipage}
\end{center}
\end{figure}
The behavior for a general value $c$ in region (1) is essentially the same. We can see that the behavior is similar to the one in Einstein gravity~\cite{Prestidge:1999uq,Hirayama:2001bi} (see Fig.\ref{t1}) except that the eigenvalue goes to negative infinity at $r_h=\sqrt{-2c}$. 
We find an unique negative mode when the heat capacity is negative. As the horizon size becomes larger, the value becomes larger and goes to zero exactly when the heat capacity changes its sign. The eigenvalue is always positive when the heat capacity is positive.

\vspace{1ex}

We show our numerical result in Fig.\ref{w2} with $(c,\Lambda)=(0.01,-1)$ in region (2).
\begin{figure}[t]
\begin{center}
\begin{minipage}{7.5cm}
\begin{center}
\includegraphics[height=7cm]{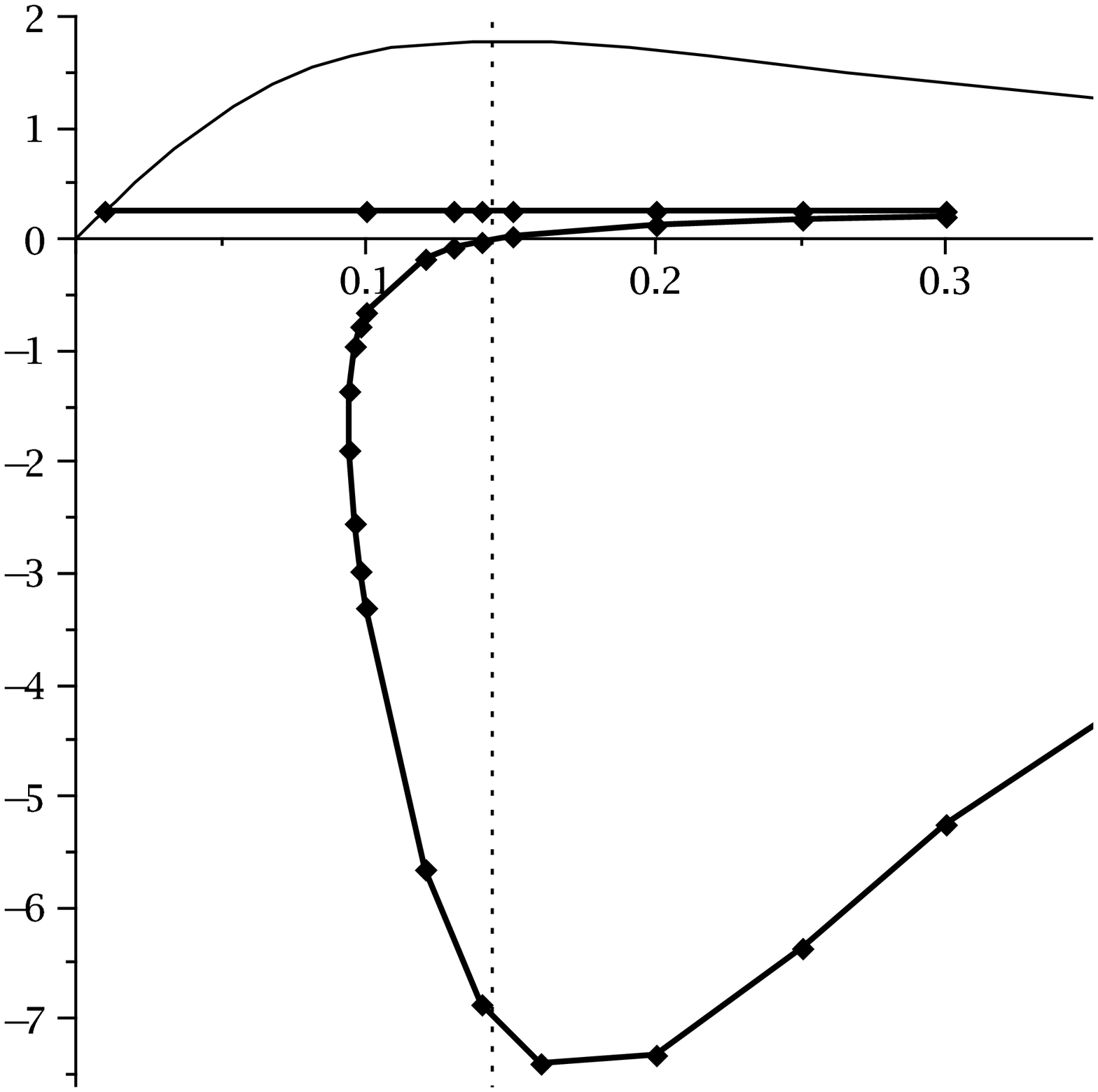}
\put(-5,144){$r_h$}
\put(-100,177){$T$}
\put(-96,20){$\lambda$}
\end{center}
\end{minipage}
\hspace{4ex}
\begin{minipage}{7.5cm}
\begin{center}
\hspace{33ex}\includegraphics[height=7cm]{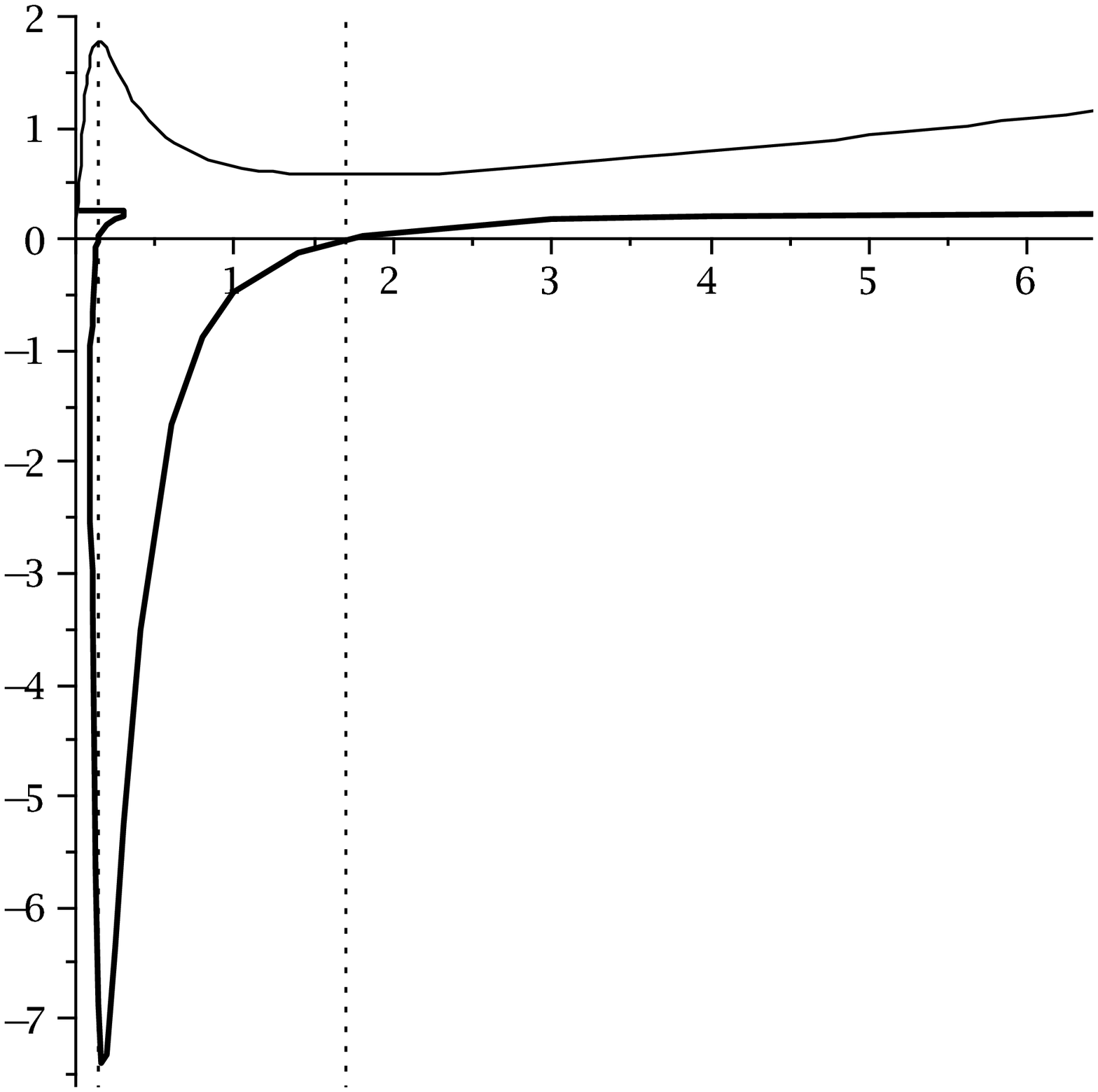}
\put(-5,144){$r_h$}
\put(-100,170){$T$}
\put(-168,20){$\lambda$}
\end{center}
\end{minipage}
\label{w2}
\caption{The eigenvalue $\lambda$ for $(c,\Lambda)=(0.01,-1)$ in terms of 
$r_h \in [0,0.4]$ (left) and $r_h \in [0,6.5]$ (right). $r_a=0.143$. $r_c=1.71$.
A small black hole $r_h\in [0.024,r_a]$ has two larger black holes with same $T$.}
\end{center}
\end{figure}
The left figure of Fig.\ref{w2} is for a small black hole where the small black hole has a positive heat capacity and thus stable as a thermodynamic system. 
A small black hole with the horizon size $r_h\in[0.024,r_a=0.143]$ has two larger black holes with a same temperature.
The behavior of eigenvalue $\lambda$ is similar to that in region (1) when a black hole has a large size. This is simply because a large black hole is weakly curved and the effect of Gauss-Bonnet term is negligible in this region.
We find no negative mode and one positive mode when a large black hole has a positive heat capacity and the eigenvalue decreases as a black hole becomes smaller and goes to zero exactly when the heat capacity changes its sign ($r_h=1.71$). 
Then the eigenvalue becomes negative when a black hole has a negative heat capacity. This behavior is same to the previous case.
However we find a new behavior in a small black hole case. As the black hole becomes more smaller, the heat capacity becomes positive again due to Gauss-Bonnet term ($r_h=0.143$). Although the negative eigenvalue quickly goes to zero, we still find negative modes until $r_h=0.094$. Moreover we find two negative modes for $r_h\in [0.094,0.143]$ and the small negative eigenvalue goes to exactly zero at $r_h=0.143$.
We also find one or two positive modes in $r_h\in [0,0.3]$.

\vspace{1ex}

We show our numerical results for $(c,\Lambda)=(1,-1)$ in Fig.\ref{w3} in region (3).
\begin{figure}[t]
\begin{center}
\begin{minipage}{7.5cm}
\begin{center}
\vspace{-3ex}
\includegraphics[height=7cm]{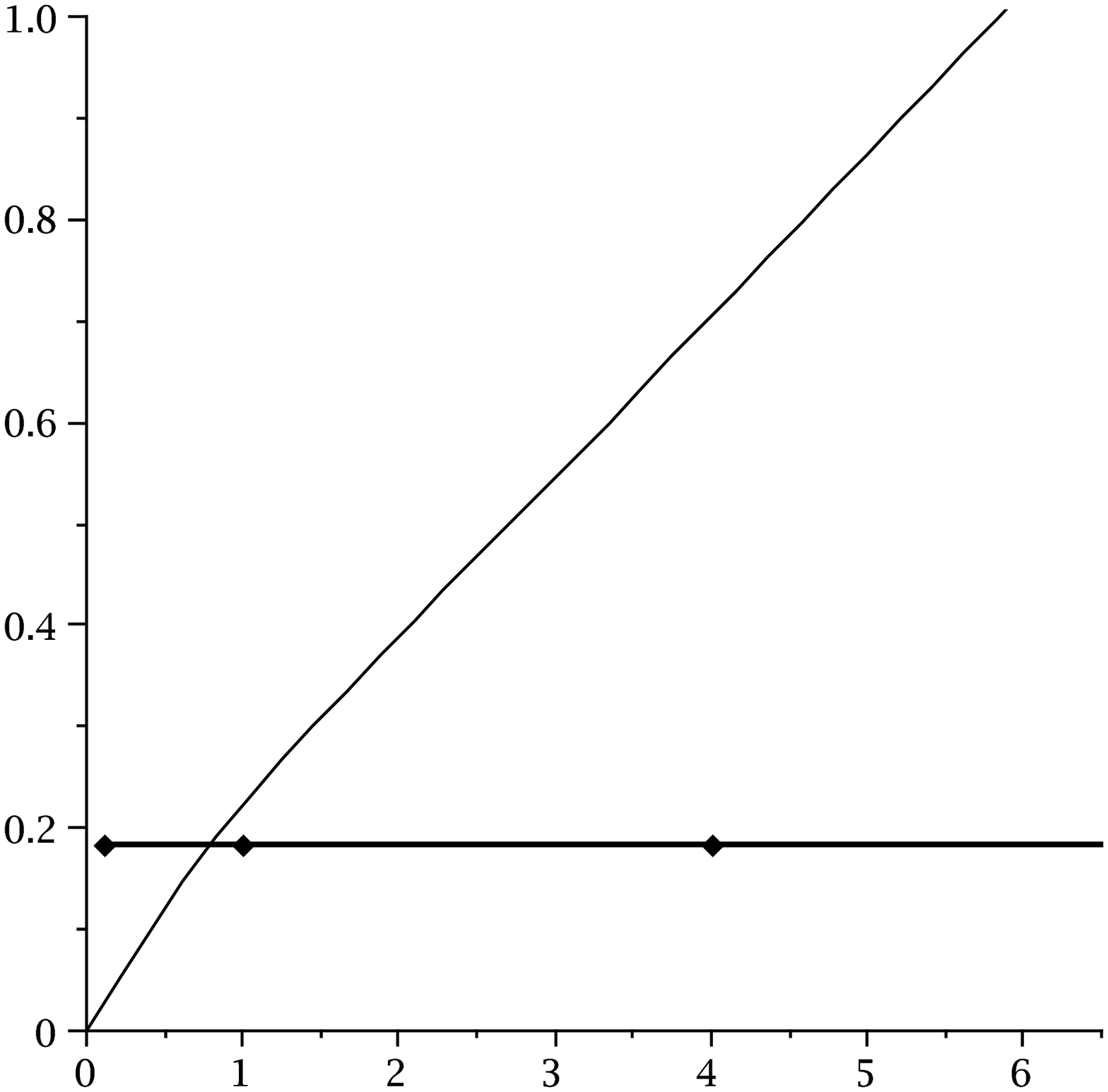}
\put(-5,4){$r_h$}
\put(-110,122){$T$}
\put(-96,30){$\lambda$}
\caption{The eigenvalue $\lambda$ in terms of $r_h$ for $c=1$ with $\Lambda=-1$.}
\label{w3}
\end{center}
\end{minipage}
\hspace{4ex}
\begin{minipage}{7.5cm}
\begin{center}\hspace*{-3ex}
\includegraphics[height=7cm]{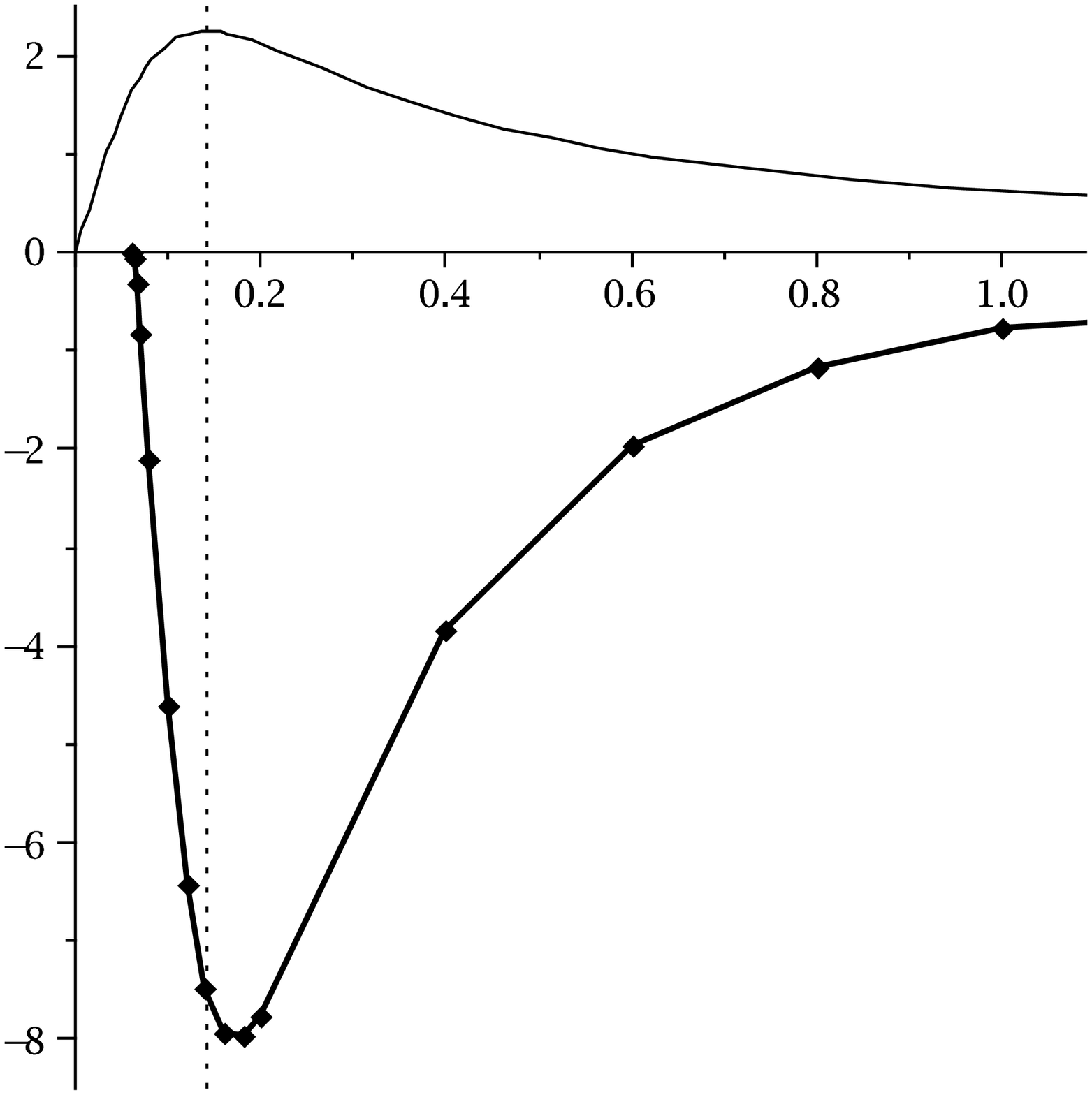}
\put(-0,150){$r_h$}
\put(-100,180){$T$}
\put(-130,30){$\lambda$}
\caption{The eigenvalue $\lambda$ in terms of $r_h \in [0,1]$ for $c=0.01$ with $\Lambda=0$. $r_c=\sqrt{2c}$.}
\label{t3}
\end{center}
\end{minipage}
\end{center}
\end{figure}
In this region the heat capacity is always positive and we find no negative mode and only find a positive mode whose eigenvalue is essentially constant and does not depend on the black hole size.

\vspace{1ex}

In the final, we show the numerical results for the case of the zero cosmological constant in Fig.\ref{t3}. We chose $c=0.01$. Notice that the equations are invariant under the rescaling \eqref{resc} with $\Lambda=0$ and we can further fix one parameter, say $c$, without losing generality.
The heat capacity is positive for small black holes ($r_h\leq\sqrt{2c}$) on the contrary to the case of Einstein gravity where a Schwarzschild black hole always has a negative heat capacity. 
We find that a negative mode still exists even a small black hole has a positive heat capacity as long as $r_h\geq 0.0612$ for $c=0.01$.

\vspace{1ex}

We now explain how the behavior of eigenmodes in region (2) continuously reaches to the behavior in region (1), (3) or the case for $\Lambda=0$, by changing $c$ or $\Lambda$.
The region (2) approaches to the region (1) by decreasing $c$ with $\Lambda$ fixed. 
Since $r_a=\sqrt{2c}+{\cal O}(c^{3/2})$ for a small $c$, the small black hole region simply disappears as $c\rightarrow 0$ and the region (2) smoothly reaches the region (1).
The region (3) can be approached by increasing $c$ from region (2). The region where the black hole has a negative heat capacity ($r_a\leq r_h\leq r_c$) shrinks as $c$ gets larger and only the most positive eigen mode remains in the region (3). 
The other two modes are getting close and become no longer normalizable modes when $c$ reaches region (3) ($c=\frac{1}{-6\Lambda}$).  We can see this tendency in Fig.\ref{w4} in Appendix~B.
The limit to $\Lambda=0$ with $c$ fixed is similarly discussed. As $\Lambda$ approaches to zero, the value of positive eigenmodes approach to zero ($\propto -\Lambda$) and these modes becomes non normalizable in the limit $\Lambda\rightarrow 0$. This is clear in Fig.\ref{t4} in Appendix~B.
In order to discuss this transition, it is more proper to introduce a finite cavity. If we introduce a finite cavity at $r=r_B$ for the case of zero cosmological constant, we have a large black hole $r_h\propto r_B$ with a same temperature and this large black hole has a positive heat capacity. Then positive eigenmodes $\lambda\propto 1/r_B^2$ are expected and become non normalizable as $r_B\rightarrow \infty$.

\vspace{1ex}

In the next section we discuss how we can understand these results. We are especially interested in the negative modes of small black holes with a positive heat capacity. It is interesting to notice that the number of negative modes in this region is two or one and there are two or one larger black holes with the same temperature for the negative or zero cosmological constant respectively.


\section{Local dynamics vs local thermodynamic vs global stability}

In region (2), we find the discrepancy between the sign of heat capacity and the existence of negative modes. This leads us to reconsider the connection between the stability against small perturbations and the local and global thermodynamic stability of black hole. 
As mentioned in Introduction, the argument on the connection between the local thermodynamic stability and the stability against small perturbations is given in detail by Reall~\cite{Reall:2001ag}, and they are shown to agree under some reasonable assumptions.

It is good to start with looking at the counter examples~\cite{Friess:2005zp} for the so-called correlated stability conjecture by Gubser and Mitra~\cite{Gubser:2000ec}. This conjecture is stated for a black string. Let us denote the direction to which a black string extends by $z$. The classical gravitational S-mode perturbations having the wavefunction $e^{ikz}$ along $z$ satisfy the Lorentzian Einstein Lichnerowicz equation \eqref{lich} of a black hole in a lower dimensions with a negative eigenvalue $\lambda=-k^2$. Gubser and Mitra conjecture that a black string with translational symmetry and infinite extent shows a Gregory-Laflamme instability if and only if it has a local thermodynamic instability. In many cases, this conjecture is checked to be true. However the counter examples are known and those black strings show a Gregory-Laflamme instability even they have the local thermodynamic stability. 
This discrepancy is understood as follow~\cite{Friess:2005zp}. The black string is not uniquely identified by the conserved charges and has moduli along which the conserved charges do not change. Since local thermodynamic of black hole cannot discuss the stability along the moduli, the Gregory-Laflamme instability can exist along the moduli and this happens in the counter examples.

Let us discuss whether a similar argument can apply for our numerical results where we found there exist two negative modes in a small black hole in region (2). First of all, there are two known black holes, $\epsilon =1$ or $-1$ in \eqref{solf}. However they have different values of the asymptotic AdS curvature in the infinity and they are not connected by a continuous parameter. 
A simple product of $n$ dimensional Schwarzschild black hole in Gauss-Bonnet theory and $S^1$ is not a solution of $n+1$ dimensional Gauss-Bonnet theory and then we cannot simply apply their argument on a black string instability here. If we can embed the Schwarzschild black hole metric in Gauss-Bonnet theory into a black string metric as a sub manifold, we may find a black string is not uniquely identified by the conserved charges. 

The non uniqueness of a black hole is discussed in~\cite{Gubser:2005ih} and the Reissner-Nordstrom black hole in four dimensions can be made classically unstable by developing a massive scalar hair under the mass and charge fixed. There is a massive scalar mode around the flat space in Gauss-Bonnet theory~\cite{Stelle:1977ry}, but the negative mode is a off-shell mode and we do not see a connection between them.

Thus it is not clear that the existence of negative modes in a small black hole has a connection to the non uniqueness of a black hole. 
On the other hand, we have an observation that having two negative modes seems to have a relation with the fact that there are two larger black holes for a smallest black hole at a same temperature, which leads us to reconsider the global stability. (The number of negative mode is one for the case of zero cosmological constant and there is one larger black hole for a small black hole at a same temperature.)

\begin{figure}[t]
\begin{center}
\includegraphics[height=7cm]{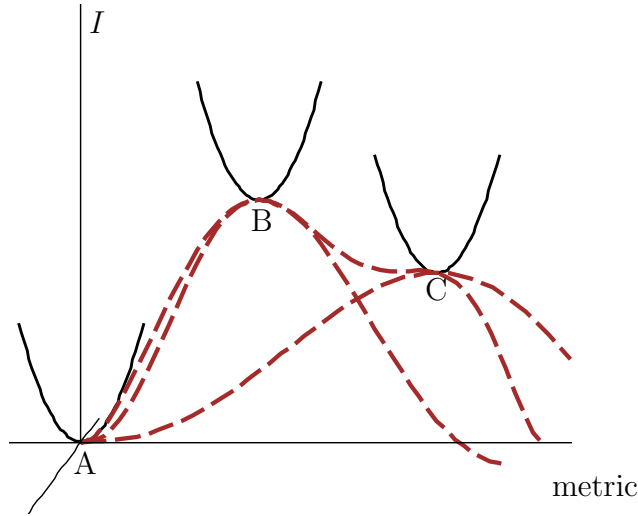}
\put(-10,10){metric}
\put(-185,185){$I$}
\put(-191,18){A}
\put(-124,110){B}
\put(-58,84){C}
\caption{The one loop effective potential around the AdS space (A),  smallest (B) and middle size (C) black holes in region (2). The tunneling processes are shown by dotted lines.}
\label{aho}
\end{center}
\end{figure}

We then refer to the original interpretation of having a negative mode as the tunneling process which induces the decay of a metastable hot flat space into nucleation of black holes in~\cite{Gross:1982cv}.  
At a temperature when the smallest black hole has two negative modes, the free energy of the smallest black hole is largest among that of three black holes and the free energy of the largest black hole is smallest.
Then we may speculate that the large negative mode indicates the tunneling process of a hot AdS space into the nucleation of the largest black holes by passing through the smallest black hole solution. 
The small negative mode may indicate a more complicated tunneling process of a hot AdS space into the nucleation of the largest black holes by first passing through the small black hole solution and next passing through a middle size black hole (Fig.\ref{aho}). 
We notice that we found positive eigenmodes around the middle size black hole.
In the entropy comparison between a black string and a black hole in the context of Gregory-Laflamme instability, the instability mode is a classical (on-shell) mode and thus we have to compare the entropy with a fixed mass when we discuss about the global thermodynamic stability. However the negative modes are off-shell modes in quantum gravity and a tunneling process is discussed at a fixed temperature.

In this section, we discussed some speculations from our numerical results, but everything is not conclusive at all.


\section{Summary and Discussions}

We studied non conformal negative modes around the Schwarzschild black hole in five dimensional Gauss-Bonnet theory with and without a negative cosmological constant. 
We found an unique negative mode when a black hole has a large radius and a negative heat capacity. The negative value goes to zero exactly when the sign of heat capacity changes its sign from negative to positive as the black hole radius becomes larger. This behavior is similar to the case of Einstein gravity and is consistent with the fact that the effect of Gauss-Bonnet term is negligible for a large black hole. On the other hand, we found negative modes still exist even when a small black hole has a positive heat capacity and the number is one or two for zero or non zero negative cosmological constant respectively. We do not fully understand why this happens and this result may indicate the black hole is not uniquely parametrized by conserved charges. Even we could refine the thermodynamic stability to be consistent with the existence of negative modes, it would be still difficult to predict the number of negative modes. 
Following the original argument in~\cite{Gross:1982cv},  we speculated that the negative modes may indicate the tunneling processes of the decay of a metastable hot AdS space into the nucleation of black holes.
However these are speculations and a more further study is necessary.

There is a report~\cite{Gleiser:2005ra} that a Schwarzschild black hole in five dimensional Gauss-Bonnet theory without the cosmological constant shows a classical instability 
against the perturbations which have arbitrary high angular momentum along $S^3$ for the range $r_h\in [0,\sqrt{2(1+\sqrt{2})c}]$ with $c>0$. This instability is expected to exist, at least, for the small negative cosmological constant ($-\Lambda r_h^2 \ll 1$). This result is also counter intuitive since an angular momentum effectively plays as a mass and we do not expect there exist an instability by heavy massive modes. Therefore the black hole thermodynamics does not capture many gravitational dynamics.

It is interesting to search a possibility to embed a Schwarzschild black hole in Gauss-Bonnet theory into a black string in a higher dimensions in order to discuss the correlated stability conjecture. In this direction it is interesting to study the Gregory-Laflamme instability for a black string solution in Gauss-Bonnet theory constructed in~\cite{Kobayashi:2004hq}.

The quantum effects become important when the spacetime curvature becomes strong and we expect a part of such quantum effects can be captured from studying the effects by higher dimensional operators like Gauss-Bonnet terms.
It would be important to understand our results in terms of quantum gravity.

\bigskip\bigskip
\subsection*{Acknowledgements}

The author would like to thank Kazuyuki Furuuchi, Gungwon Kang, Makoto Narita and Dan Tomino. 
He would like to thank Korea Institute for Advanced Study for a warm hospitality during his visit where the work was finished.
This work is supported by National Center for Theoretical Sciences, Taiwan,
(No. NSC 97-2119-M-002-001).

\appendix

\section{Numerical analysis}

We explain the procedure of numerical analysis. We solve the equations of motion
\begin{align}
 \Delta_{~b~d}^{a~c}(\bar{g}) \delta g_c^{~d} &= \lambda \delta g^a_{~b} ,
\end{align}
with the ansatz
\begin{align}
  \delta g_a^{~b} = & {\rm diag} ( H_{tt}(r), H_{rr}(r), K(r), K(r), K(r) ) .
\end{align}
We can see most of the equations are solved with the above ansatz and finally have three nontrivial equations which is schematically written
\begin{align}
  H_{tt}'(r) &= F_1( H_{tt}(r), H_{rr}(r), K(r) ) ,
  \\
  H_{rr}'(r) &= F_2( H_{tt}(r), H_{rr}(r), K(r) ) ,
  \\
  K''(r) &= F_3(K'(r), H_{tt}(r), H_{rr}(r), K(r) ) ,
\end{align}
where $F_i$ ($i=1,2,3$) are functions of variables written inside the bracket and $H_{tt}'(r)=d H_{tt}(r)/d r$ etc. From counting the number of variables and equations, it is easy to see that there are two independent modes in these perturbations which we refer the 'traceless' and 'trace' modes since they are traceless and trace modes in Einstein gravity limit.

We solve the equations of motion from the horizon to the infinity. Because of the invariance under the rescaling \eqref{resc}, we can take $\Lambda=-1$ in the numerical analysis. The parameters are $c$, $r_h$ and $\lambda$. We fix $c$ and $r_h$ and change $\lambda$ to find eigenmodes by the shooting method.

We analytically solve the equations of motion near the horizon up to $\epsilon^4$, ($\epsilon=r-r_h$), for $H_{tt}$ and $H_{rr}$ and up to $\epsilon^5$ for $K(r)$ and compute the boundary condition at a small value $\epsilon$, i.e. $H_{tt}(r_h+\epsilon)$, $H_{rr}(r_h+\epsilon)$, $K(r_h+\epsilon)$ and $K'(r_h+\epsilon)$. We took $\epsilon=0.001$ for the cases $\Lambda=-1$, $c=-0.1,0,0.01,1$ and $r_h\in [0,10]$ in Fig.\ref{w1}-\ref{w3}. We then solve the equations of motion numerically using a Fehlberg fourth-fifth order Runge-Kutta method with degree four interpolant in Maple 11 up to a large value $r_f$ in $r$, $r_f=10^3$ in Fig.\ref{w1}-\ref{w3}.

We then fit the numerical values $H_{tt}(r_f)$, $H_{rr}(r_f)$, $K(r_f)$ and $K'(r_f)$ by the asymptotic solutions at infinity. We analytically compute the asymptotic solutions up to the second leading orders in $r$. There are four solutions since there are normalizable and non-normalizable solutions of 'traceless' and 'trace' modes. We fit the numerical values by using these all four solutions, since even we start the boundary condition near the horizon such that the 'trace' mode is set to be zero, there always exist numerical errors and the 'trace' modes are included in the numerical values. We them obtain the coefficient of the non-normalizable solution of 'traceless' mode in the numerical values and study when the coefficient becomes zero by changing $\lambda$ using the shooting method.

We checked our numerical analysis has a good accuracy from various approaches. Our results approach to the one in Einstein gravity by taking $c$ smaller values. We change the value of $\Lambda$ and checked the invariance under the rescaling is satisfied with a good accuracy. We changed the value of $\epsilon$ and $r_f$ and studied the results do not change.

In the case of zero cosmological constant, the procedure is same except that $r_f=100$. Since the asymptotic solutions are exponential ($\propto e^{\pm\sqrt{-2\lambda}r}$) for the 'traceless' mode and sine ($\propto \sin(\sqrt{\pm2\lambda/3})$) for the 'trace' mode, the exponential growing is already manifest at $r_f=100$.


\section{Numerical results}

\begin{figure}[h]
\begin{center}
\begin{minipage}{7.5cm}
\begin{center}
\hspace*{-4ex}
\includegraphics[height=7cm]{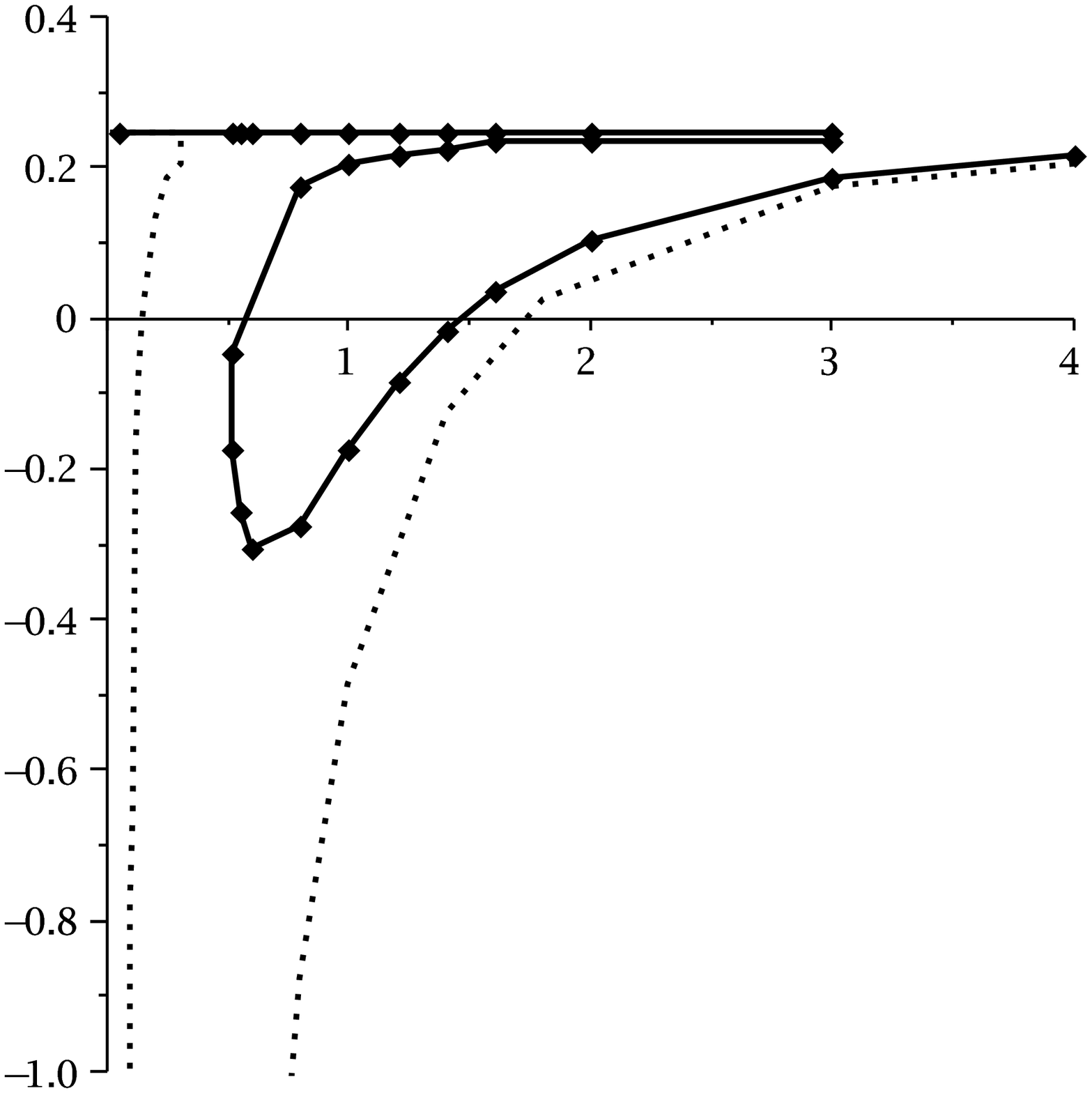}
\put(-5,140){$r_h$}
\put(-160,30){$\lambda$}
\caption{$\lambda$ for $(c,\Lambda)=(0.1,-1)$. The dotted line is $\lambda$ for $(c,\Lambda)=(0.01,-1)$.}
\label{w4}
\end{center}
\end{minipage}
\hspace{4ex}
\begin{minipage}{7.5cm}
\begin{center}\hspace*{-3ex}
\includegraphics[height=7cm]{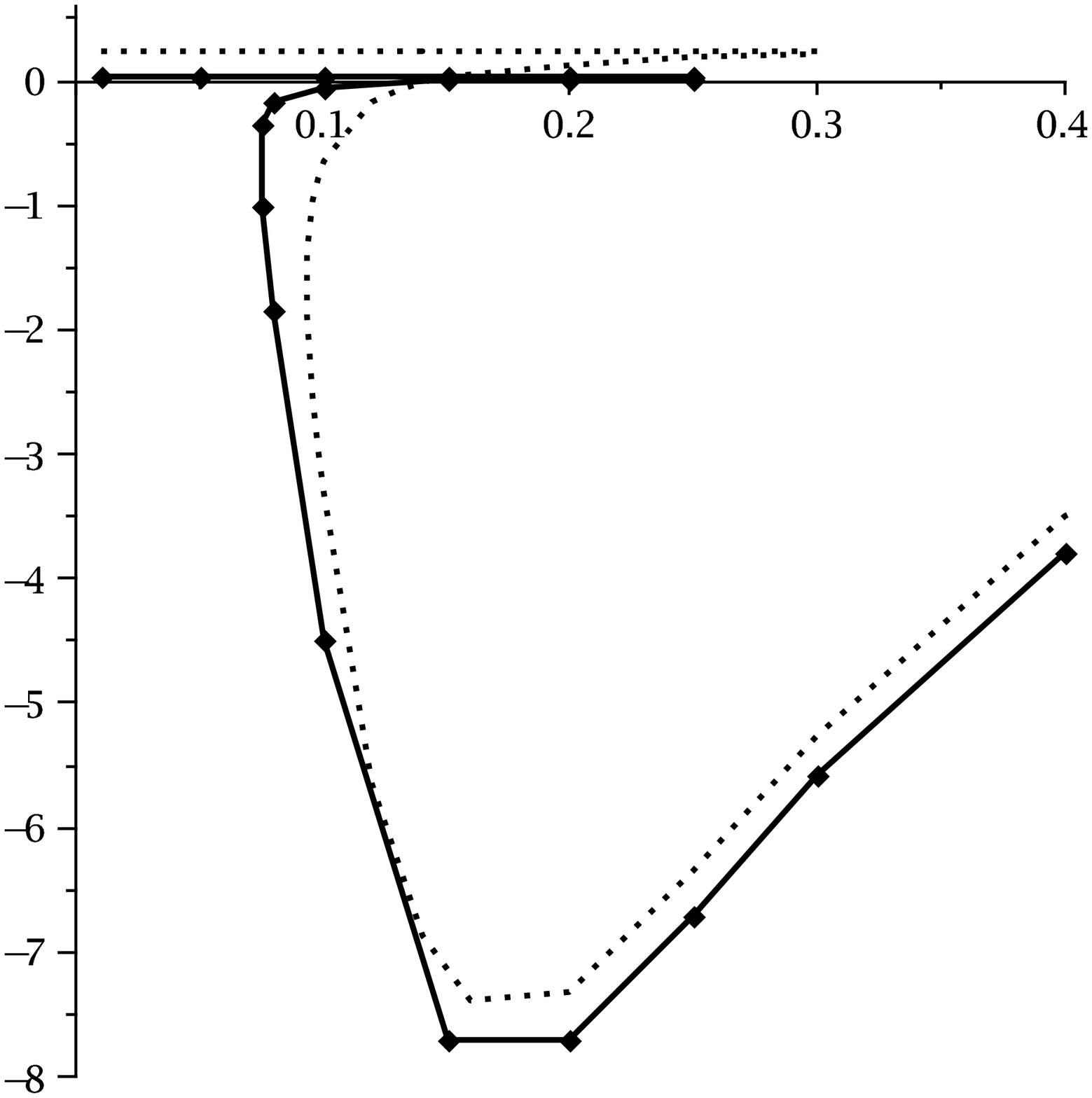}
\put(-5,185){$r_h$}
\put(-170,30){$\lambda$}
\caption{$\lambda$ for $(c,\Lambda)=(0.01,-0.1)$.  The dotted line is $\lambda$ for $(c,\Lambda)=(0.01,-1)$.}
\label{t4}
\end{center}
\end{minipage}
\end{center}
\end{figure}

We show our numerical results. These values are used to plot the figures. We employ the shooting method and change $\lambda$ by $0.01$ to find the eigenvalue. Thus for example the first value $(0.8, -2.925)$ means the real value $\lambda$ is located between $-2.92$ and $-2.93$.
\begin{itemize}
 \item[(1)]
	   {\boldmath $(c, \Lambda)=(-0.1,-1)$} : Fig.\ref{w1}

\hspace*{-6ex}
$\begin{array}{|c|c|c|c|c|c|c|c|c|c|c|c|}
  \hline
 r_h         & 0.8    &0.9     &1         &1.2     &1.4    &1.6    &1.8     &2         &3        &4        
 &10
 \\
  \hline
 \lambda&-2.925&-1.795 &-1.185 &-0.585 &-0.305&-0.145&-0.045&0.015&0.155&0.205
 &0.245
 \\
  \hline
 \end{array}$
 \item[(2)]
	   {\boldmath $(c, \Lambda)=(0.01,-1)$} : Fig.\ref{w2}

\hspace*{-6ex}
$\begin{array}{|c|c|c|c|c|c|c|c|c|c|c|c|}
 \hline
 r_h         &0.01&0.1&0.13&0.14&0.15&0.2&0.25&0.3
 \\ \hline
 \lambda&0.245&0.245&0.245&0.245&0.245&0.245&0.245&0.245
  \\ \hline
 \end{array}$

\hspace*{-6ex}
$\begin{array}{|c|c|c|c|c|c|c|c|c|c|c|c|}
 \hline
 r_h         &0.3&0.25&0.2&0.15
 \\ \hline
 \lambda&0.205&0.185&0.135&0.025
  \\ \hline
 \end{array}$

\hspace*{-6ex}
$\begin{array}{|c|c|c|c|c|c|c|c|c|c|c|c|}
 \hline
 r_h         &0.14&0.13&0.12&0.1&0.098&0.096&0.094
 \\ \hline
 \lambda&-0.015&-0.075&-0.175&-0.655&-0.775&-0.965&-1.365
  \\ \hline
 \end{array}$

\hspace*{-6ex}
$\begin{array}{|c|c|c|c|c|c|c|c|c|c|c|c|c|}
 \hline
 r_h         &0.094&0.096&0.098&0.1&0.12&0.14&0.16&0.2&0.25
 \\ \hline
 \lambda&-1.885&-2.545&-2.965&-3.315&-5.645&-6.865&-7.395&-7.315&-6.355
  \\ \hline
 \end{array}$

\hspace*{-6ex}
$\begin{array}{|c|c|c|c|c|c|c|c|c|c|c|c|c|}
 \hline
 r_h         &0.3&0.4&0.6&0.8&1&1.4&1.8&3&4&10
 \\ \hline
 \lambda&-5.255&-3.505&-1.665&-0.875&-0.485&-0.125&0.025&0.175&0.205&0.245
  \\ \hline
 \end{array}$

 \item[(3)]
	   {\boldmath $(c, \Lambda)=(1,-1)$} : Fig.\ref{w3}

\hspace*{-6ex}
$\begin{array}{|c|c|c|c|c|c|c|c|c|c|c|c|c|}
 \hline
 r_h         &0.1&1&4&10
 \\ \hline
 \lambda&0.185&0.185&0.185&0.185
  \\ \hline
 \end{array}$


 \item[(4)]
 	   {\boldmath $(c, \Lambda)=(0.01,0)$} : Fig.\ref{t3}

\hspace*{-6ex}
$\begin{array}{|c|c|c|c|c|c|c|c|c|c|c|c|c|}
 \hline
 r_h         &0.0612&0.062&0.065&0.07&0.08&0.1&0.12&0.14&0.16
 \\ \hline
 \lambda&-0.015&-0.065&-0.315&-0.835&-2.095&-4.595&-6.415&-7.465&-7.925
  \\ \hline
 \end{array}$

\hspace*{-6ex}
$\begin{array}{|c|c|c|c|c|c|c|c|c|c|c|c|c|}
 \hline
 r_h         &0.18&0.2&0.4&0.6&0.8&1&2&4&10
 \\ \hline
 \lambda&-7.965&-7.765&-3.835&-1.975&-1.175&-0.765&-0.195&-0.055&-0.0085
  \\ \hline
 \end{array}$


\item[(5)]
 	   {\boldmath $(c, \Lambda)=(0.1,-1)$} : Fig.\ref{w4}

\hspace*{-6ex}
$\begin{array}{|c|c|c|c|c|c|c|c|c|c|c|c|c|}
 \hline
 r_h         &0.05 &0.52 &0.55 &0.6 &0.8 &1 &1.2 &1.4 &1.6 &2.0 &3.0
 \\ \hline
 \lambda&0.245 &0.245 &0.245 &0.245 &0.245 &0.245 &0.245 &0.245 &0.245 &0.245 &0.245
  \\ \hline
 \end{array}$

\hspace*{-6ex}
$\begin{array}{|c|c|c|c|c|c|c|c|c|c|c|c|c|}
 \hline
 r_h         &3.0 &2.0 & 1.6 &1.4 &1.2 &1 &0.8 &0.52
 \\ \hline
 \lambda&0.235 &0.235 &0.235 &0.225 &0.215 &0.205&0.175 &0.045
  \\ \hline
 \end{array}$

\hspace*{-6ex}
$\begin{array}{|c|c|c|c|c|c|c|c|c|c|c|c|c|}
 \hline
 r_h         &0.52 &0.55 &0.6 &0.8 &1 &1.2 &1.4 &1.6 &2.0 &3.0& 4.0
 \\ \hline
 \lambda&-0.175&-0.255&-0.305&-0.275&-0.175&-0.085&-0.015&0.035&0.105&0.185&0.215
  \\ \hline
 \end{array}$




\item[(6)]
 	   {\boldmath $(c, \Lambda)=(0.01,-0.1)$} : Fig.\ref{t4}

\hspace*{-6ex}
$\begin{array}{|c|c|c|c|c|c|c|c|c|c|c|c|c|}
 \hline
 r_h         &0.01 &0.05 &0.1 &0.15 &0.2 &0.25
 \\ \hline
 \lambda&0.025&0.025&0.025&0.025&0.025&0.025
  \\ \hline
 \end{array}$

\hspace*{-6ex}
$\begin{array}{|c|c|c|c|c|c|c|c|c|c|c|c|c|}
 \hline
 r_h         &0.25 &0.2 &0.15 &0.1 &0.08 &0.075
 \\ \hline
 \lambda&0.015&0.015&0.005&-0.045&-0.175&-0.355
  \\ \hline
 \end{array}$

\hspace*{-6ex}
$\begin{array}{|c|c|c|c|c|c|c|c|c|c|c|c|c|}
 \hline
 r_h         &0.075 &0.08 &0.1 &0.15 &0.2 &0.25 &0.3 &0.4
 \\ \hline
 \lambda&-1.005 &-1.835&-4.485&-7.705&-7.715&-6.705&-5.575&-3.795
  \\ \hline
 \end{array}$




\end{itemize}


\bigskip\bigskip

\end{document}